\documentclass[preprint,showpacs,preprintnumbers,amsmath,amssymb]{revtex4}

\usepackage{graphicx}
\usepackage{dcolumn}
\usepackage{bm}
\usepackage{color}

\begin{document}

\preprint{Submitted July, 2008}

\newcommand{\comm}[1]{\textcolor{blue}{(#1)}}

\title{A New Car-Following Model Inspired by Galton Board}

\author{Fa Wang, Li Li, Xuexiang Jin, Jianming Hu, Yi Zhang, Yan Ji}
 \altaffiliation{Department of Automation, Tsinghua University, Beijing 100084, P. R. China}
 \email{li-li@mail.tsinghua.edu.cn}

\date{\today}

\begin{abstract}
Different from previous models based on scatter theory and random
matrix theory, a new interpretation of the observed log-normal type
time-headway distribution of vehicles is presented in this paper.
Inspired by the well known Galton Board, this model views driver's
velocity adjusting process similar to the dynamics of a particle
falling down a board and being deviated at decision points. A new
car-following model based on this idea is proposed to reproduce the
observed traffic flow phenomena. The agreement between the empirical
observations and the simulation results suggests the soundness of
this new approach.
\end{abstract}


\maketitle

\section{Introduction}
\label{sec:1}

To explain and reproduce the complex phenomena of road traffic, the
dynamics of traffic flows are often described on $N$ strongly-linked
particles (vehicles) under fluctuations
\cite{ChowdhurySantenSchadschneider2000}, \cite{Helbing2001},
\cite{MahnkeKaupuzsLubashevsky2005}. Since the governing interaction
forces or potentials cannot be directly measured, the statistical
distributions of particles are often investigated instead
\cite{Kerner2004}, \cite{KernerKlenovHillerRehborn2006}.

Among different statistical features, the distributions of
space-gaps/time-headways between these particles (vehicles) received
consistent interests from various viewpoints. In transportation
engineering, the corresponding investigation data are helpful not
only in understanding the microscopic-level driving behaviors but
also in estimating the macroscopic-level roadway capacities
\cite{Luttinen1992}, \cite{MichaelLeemingDwyer2000}, \cite{HCM2000},
\cite{ZhangWangWeiChen2007}. Numerous distribution models have been
developed over the past 50 years to directly fit the empirical data.
In some recent studies \cite{KrbalekSebaWagner2001},
\cite{KrbalekHelbing2004}, \cite{HelbingTreiberKesting2006},
\cite{Abul-Magd2007}, some theoretical models were presented from
various physical perspectives (e.g. scatter theory and random matrix
theory) to explain why we can observe similar distributions even for
different phases (i.e. Kerner's the free-flow, synchronized flow,
and moving jam phases \cite{Kerner2004}).

Differently in this paper, another interpretation inspired by the
well-known Galton board is presented. A new car-following model
based on this interpretation is also proposed to reproduce the
observed traffic flow phenomena. The research purposes are twofold
here: 1) Previous physical interpretations focus on the steady-state
macroscopic-level statistics; while this method provides a
microscopic-level dynamic explanation, which can also be used to
simulation the transient-state statistics of inter-arrival and
inter-departure vehicle queuing interactions. 2) It is interesting
that as a slightly modified extension of the classic car-following
model, this new model can easily reproduce the observed time-headway
distributions, which had been neglected by many previous approaches.

\section{The Log-Normal Distribution Model of Time-Headways}
\label{sec:2}

\begin{figure}[h]
  \resizebox{0.75\columnwidth}{!}{%
    \includegraphics{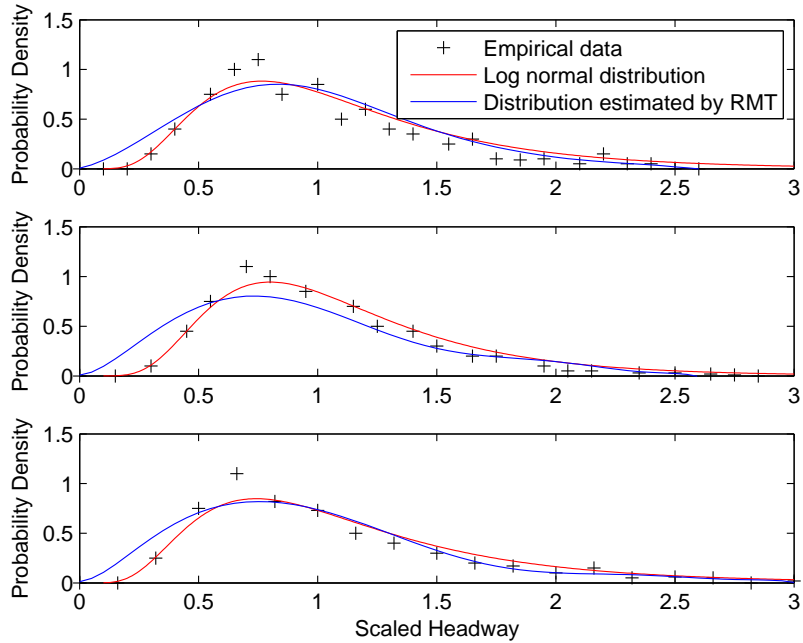}
  }
  \caption{Probability density $P(\tau)$ for scaled empirical time headways $\tau$
  between successive cars in traffic flow. For comparison, the empirical data and
  super-statistical distributions on random matrix theory are
  excerpted from \cite{KernerKlenovHillerRehborn2006}.}
\label{fig:1}       
\end{figure}

Based on several recent studies
\cite{KernerKlenovHillerRehborn2006}, \cite{ZhangWangWeiChen2007},
\cite{ThiemannTreiberKesting2008}, \cite{KrbalekSebaWagner2001},
\cite{Abul-Magd2007}, we believe that log-normal distribution model
is a simple yet effective model comparing to other ones, i.e.
distribution models according to random-matrix theory
\cite{KernerKlenovHillerRehborn2006}, \cite{Abul-Magd2007}. For
example, Fig.~\ref{fig:1} shows the comparison results for the
empirical, super-statistical on random matrix theory
\cite{Abul-Magd2007} and log-normal distributions $P(\tau)$, where
$\tau = t_h / \langle t_h \rangle$, $t_h$ denotes the sampled time
headways. It is clear that log-normal distribution model leads to
smaller fitting errors.

%

Moreover, log-normal distribution model can also be used to depict
the distributions of time headways in transient-state flow. For
instance, the departure headways (which are usually defined as the
times that elapse between consecutive vehicles when vehicles in a
queue start crossing the stop line) are shown to follow log-normal
distributions in \cite{SuWeiChengYaoZhangLiZhangLi2008}.

Noticing that the random matrix theory gives a soundable explanation
for its corresponding distribution model, an interesting question
naturally arises as: ``Can we provide a physical interpretation of
such log-normal distributions?'' After comparing with the other
dynamic processes that yield log-normal distribution
\cite{Galton1889}, \cite{AitchisonBrown1957},
\cite{CrowShimizu1988}, \cite{LimpertStahelAbbt2001}, we think the
outcomes of such distributions can be explained as follows.

In car-following process, the driver of the following vehicle will
adjust his/her velocity from time to time to track the leading
vehicle and meanwhile keep a safe distance between the leading
vehicle and him/her. Because the leading vehicle's movement is often
unpredictable (at least not fully predictable), the accelerating and
braking action of the driver is often overdue.

\begin{figure}[h]
  \resizebox{0.45\columnwidth}{!}{%
    \includegraphics{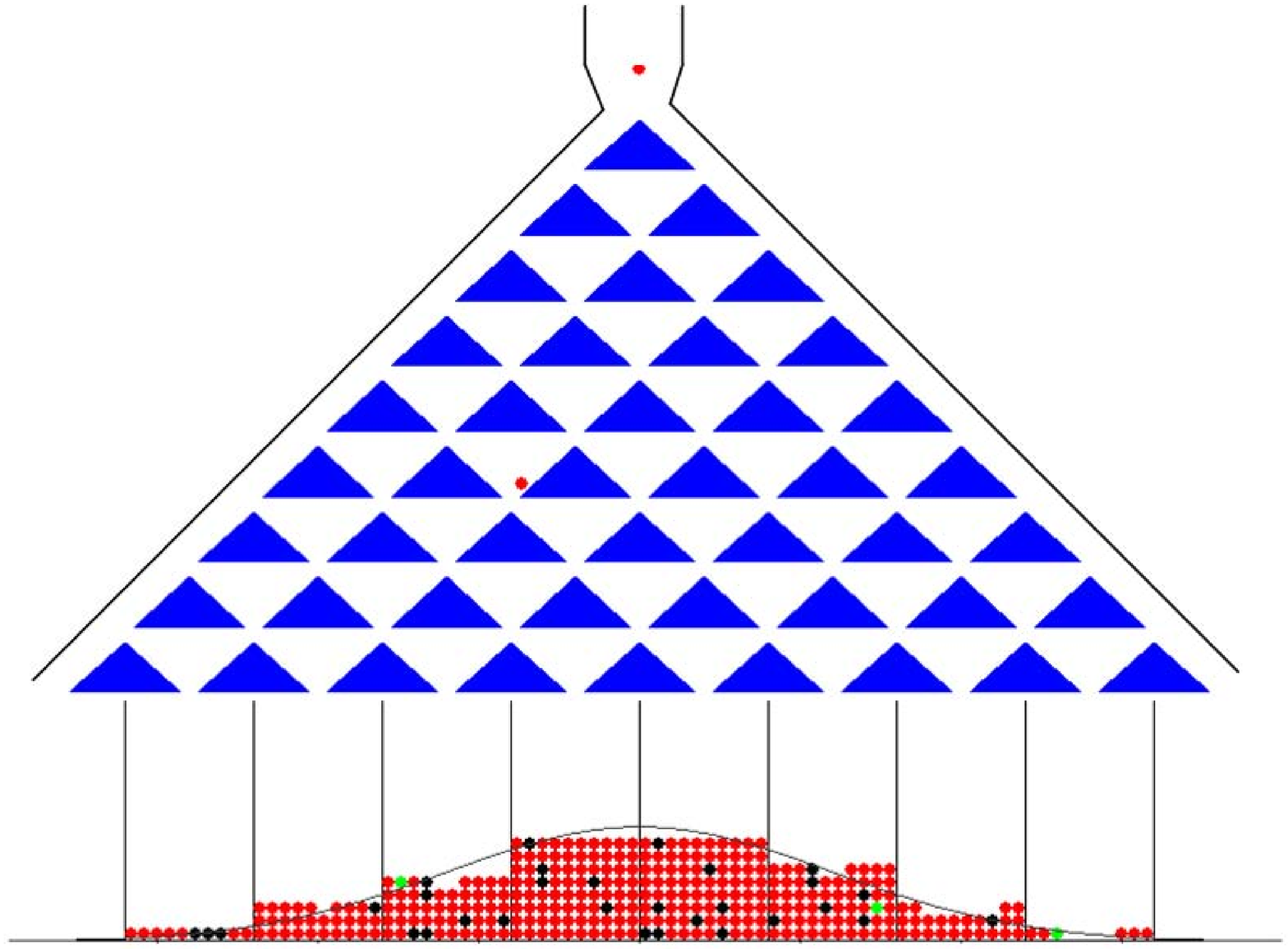}
  }
  \centerline{\footnotesize (a)}
  \resizebox{0.5\columnwidth}{!}{%
    \includegraphics{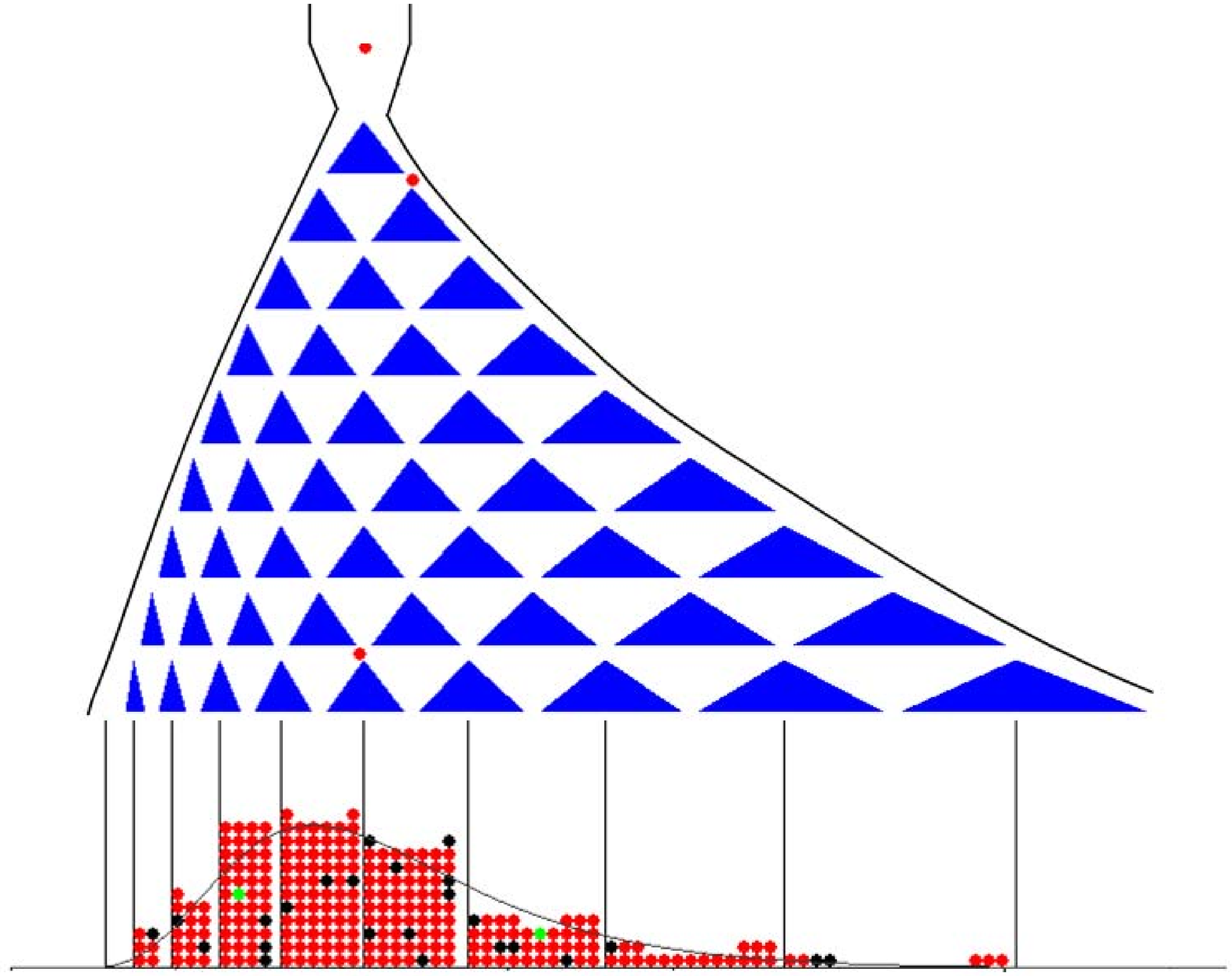}
  }
  \centerline{\footnotesize (b)}
  \caption{Diagram of two Galton boards yielding normal (a) and log-normal
  (b) distributions respectively. If the tip of a triangle is at distance $x$
  from the left edge of the board, triangle tips to the right and to the left
  below it are placed at $x+c$ and $x-c$ for the normal distribution, and
  $x \cdot c'$ and $x / c'$ for the log-normal, where $c$ and $c'$ are
  constants.}
\label{fig:2}       
\end{figure}

This persistent velocity adjusting process is somewhat like the
process of the particles falling down a board and being deviated at
decision points (the tips of the triangular obstacles) either left
or right with equal probability, see Fig.~\ref{fig:2}. If the
deviation of the particle from one row to the next is a random
additive process with possible values $+c$ and $-c$, the normal
distribution will be created by the board, which reflects the
cumulative additive effects of the sequence of decision points. But
if the deviation of the particle from one row to the next is a
random multiplicative process with possible values $\cdot c'$ and $/
c'$, the log-normal distribution will be generated.

%

It is obvious that drivers tend to take very careful acceleration
when the spacing between the two vehicles is small and do not want
to speed up at once after braking. Thus, the deviation of the
particle to the left is equivalent to the following vehicle's
accelerating action, since the relative speed decrease slowly at
this period; while the deviation to the right is equivalent to the
following vehicle's braking action, since the relative speed
increase quickly at this period. From this view point, the
car-following process is similar to the particles falling in the
log-normal type Galton board. To verify this conjecture, a
microscopic simulation model inspired by log-normal type Galton
board is proposed in the next section to reproduce the observed
phenomena.

\section{The New Car-Following Model Inspired by Galton Board}
\label{sec:3}

In this section, it will be interesting to find that a car-following
model incorporating Galton Board mechanism can be easily derived
from a famous classic car-following model proposed in later 1950s
\cite{GazisHermanPotts1959}, \cite{GartnerMesserRathi1997}. The
proposed model mainly depicts three states of a vehicle: 1) stopped,
2) starting-up, and 3) driving/braking states. More precisely, they
can be described as follows.

Suppose $v_i(t)$ and $x_i(t)$ are the velocity and position of the
$i$th vehicle (follower) at time $t$, respectively. Similarly,
$v_{i-1}(t)$ and $x_{i-1}(t)$ are the velocity and position of the
${i-1}$th vehicle (leader) at time $t$. $g_i(t) = x_i(t) -
x_{i-1}(t) - l_i$ denotes the gap between the two vehicles at time
$t$. $T \ge 0$ is the simulation time span.

\indent

1) If $v_i(t) = 0$, that is, the $i$th vehicle is fully stopped at
time $t$; it will stay stopped in the next time interval $T$ and
then enter the starting state since time $(t+T)$.

2) Else if the $i$th vehicle is in the starting-up state at time
$t$, it will check its coming actions from the following three
choices:

\indent \indent 2.a) If $g_i(t) < G_{min}$, where $G_{min}$ is the
minimum stop distance. The $i$th vehicle will go back to stop state
by letting $v_i(t+T) = 0$ without doing anything else (that's,
$x_i(t+T) = x_i(t)$.

\indent \indent 2.b) Else if $v_i(t) < v_{starting}$, where
$v_{starting}$ is the limit of the staring-up velocity; the vehicle
is still in the starting-up state at time $(t+T)$. Thus, it will
accelerate with the staring-up rate $a_{starting}$ in the next time
interval $T$ till time $(t+T)$.

\indent \indent 2.c) Otherwise, the vehicle will switch to the
driving/braking state since time $t$ and re-determine its action
according to the updating rules set for the driving/braking state at
time $t$.

3) Otherwise, the vehicle is running in the driving/braking state.
Actually, the driving/braking state contains three modes: 3.a)
free-driving mode, 3.b) braking model and 3.c) following mode (see
Fig.~\ref{fig:3}), which can be represented in details as:

%
\begin{figure}[h]
\center
\resizebox{0.5\columnwidth}{!}{%
  \includegraphics{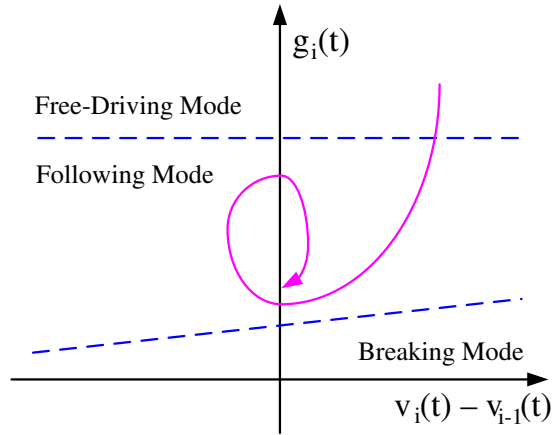}
}
\caption{Diagram of the three modes (regions) in the gap vs.
relative-speed space. Comparing to Fritzsche 1994 model used in
PARAMICS and Wiedemann Psychophysical model used in VISSIM, our
model has fewer modes.}
\label{fig:3}       
\end{figure}

\indent \indent 3.a) If $g_i(t) > G_{max}$, where $G_{max}$ is the
maximum coupling distance; the $i$th vehicle is in free-driving
mode. It will try to approach the highest velocity as

\begin{equation}
\label{equ:1} v_{i}(t+T) = \min \{v_{i}(t) + a_{max}^+ \times T,
v_{max} \}
\end{equation}

\noindent where $a_{max}$ is the maximum allowable acceleration
rate. For simplicity, we choose $T=1$ second here.

\indent \indent 3.b) Else if $v_i(t) - v_{i-1}(t) > [g_i(t) - G] /
H$, there exists a risk to collide, and $i$th vehicle is in braking
mode. It will try to stop as quickly as possible

\begin{equation}
\label{equ:2} v_{i}(t+T) = v_{i}(t) - D \times T
\end{equation}

\noindent where $G$, $H$ and $D$ are positive constants denoting the
minimum safety gap, deceleration time and braking decelerating rate,
respectively.

\indent \indent 3.c) Otherwise, the $i$th vehicle is in the
following mode. We have the following two-step updating rules as

\begin{equation}
\label{equ:3} \widetilde{v}_i(t+T) = \left\{
\begin{array}{ll} \max \{ \beta v_i(t) \frac{x_i(t) - x_{i-1}(t) - l_i}{x_i(t-T)
- x_{i-1}(t-T) - l_i}, 0 \}, \indent \textrm{with probability $p$ to decelerate}\\
\min \{ \frac{1}{\beta} v_i(t) \frac{x_i(t) - x_{i-1}(t) -
l_i}{x_i(t-T) - x_{i-1}(t-T) - l_i}, v_{max} \}, \indent
\textrm{with probability $(1-p)$ to accelerate}
\end{array} \right.
\end{equation}

\begin{equation}
\label{equ:4} v_i(t+T) = \left\{
\begin{array}{ll} \max \{ \widetilde{v}_i(t), v_{i-1}(t) - a_{max}^- \times T\},
\indent \textrm{if decelerate} \\
\min \{ \widetilde{v}_i(t), v_{i-1}(t) + a_{max}^+ \times T\},
\indent \textrm{if accelerate}
\end{array} \right.
\end{equation}

\noindent where $l_i$ denotes the length of the $i$th vehicle.
$v_{max}$ is the maximum allowable velocity. $a_{max}^+$ and
$a_{max}^-$ are the maximum allowable car-following ac/deceleration
rates. The $\max$, $\min$ functions are added to guarantee that the
velocity and ac/decelerating rates are within the limits. $\beta$ is
a positive constant to be determined according to investigation
data.

Eq.(\ref{equ:3})-(\ref{equ:4}) characterize the simulated time
headways to follow the log-normal distributions. A brief explanation
of this mechanism is provided in the Appendix. It is assumed that $0
< \beta \lesssim 1$ together with $0 < p < 0.5$, since drivers tend
to keep a close spacing. We can see that
Eq.(\ref{equ:3})-(\ref{equ:4}) naturally incorporate the
frequently-used randomly-slow-down features of driving behaviors.

4) Finally, the position of vehicle is updated as

\begin{equation}
\label{equ:5} x_i(t + T) = x_i(t) + v_i(t+T) \times T
\end{equation}

\indent

To illustrate the effectiveness of the proposed model, a specialized
form is used to reproduce the complex phenomena of freeway traffic
flows. Particularly, we have all the vehicles have the same length
as $l_i = 4$m and $T = 1$s. Other parameters are set as
$v_{starting} = 8$m/s, $a_{starting} = 4$m/s$^2$, $G_{min} = 0.5$m,
$G_{max} = 52.5$m, $v_{max} = 30$m/s, $a_{max}^+ = 4$m/s, $a_{max}^-
= 8$m/s, $\beta = 0.93$, $p = 0.3$. $G=0.5$m, $H=7$s, $D=8$m/s$^2$.

\begin{figure}[h]
  \resizebox{0.6\columnwidth}{!}{%
  \includegraphics{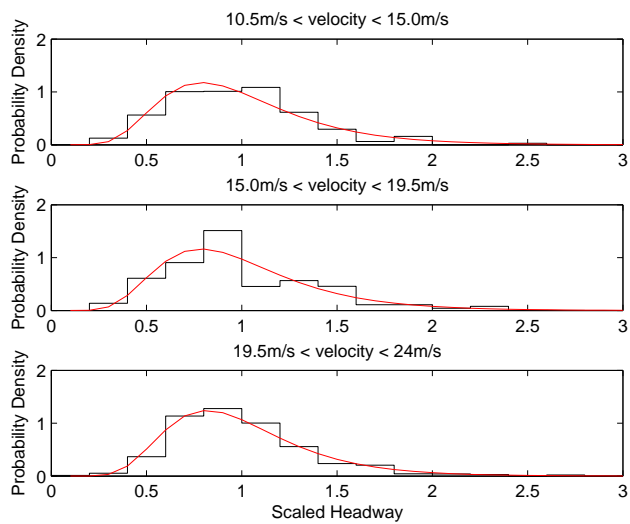}
  }
  \centerline{\footnotesize (a)}
  \resizebox{0.6\columnwidth}{!}{%
    \includegraphics{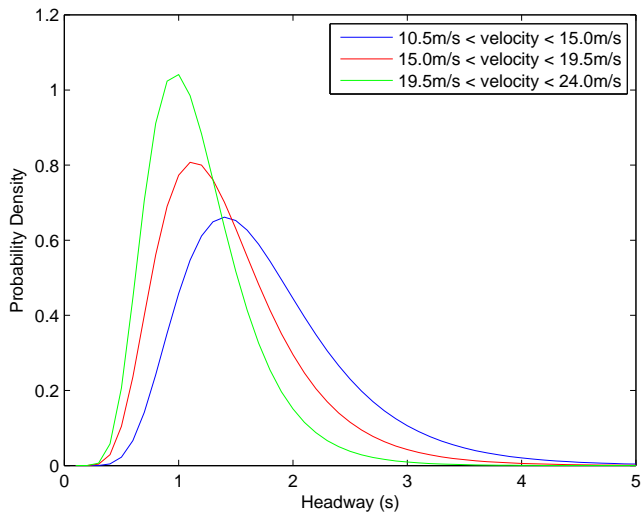}
  }
  \centerline{\footnotesize (b)}
  \caption{Probability density for (a) scaled simulated time headways $\tau$ and
  (b) un-scaled simulated time headways $t_h$ between successive cars in traffic flow.}
\label{fig:4}       
\end{figure}

Fig.~\ref{fig:4} shows the corresponding distributions of the
simulated time headways within different velocity ranges. The
log-normal distribution model passes the Kolmogorov-Smirnov
(K-S)hypothesis test. This proves that the new model yields the
log-normal type time-headway distribution as observed; while many
previous models yield symmetric distributions, i.e. Fig.~7 of
\cite{Bando1995}. Moreover, as the velocity of the vehicle
increases, the mean time headways of such log-normal type
distributions will approach the saturation headway. This fits the
observations \cite{KernerKlenovHillerRehborn2006},
\cite{SchonhofHelbing2007}, too.

%

\begin{figure}[h]
  \resizebox{0.6\columnwidth}{!}{%
  \includegraphics{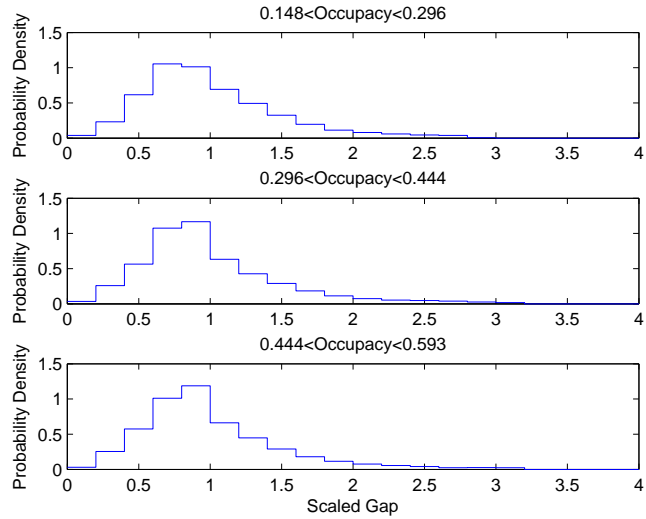}
  }
  \centerline{\footnotesize (a)}
  \resizebox{0.6\columnwidth}{!}{%
    \includegraphics{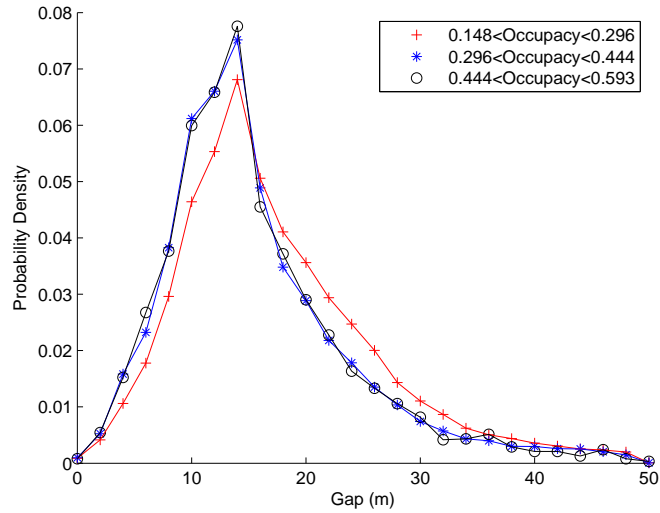}
  }
  \centerline{\footnotesize (b)}
  \caption{Probability density for (a) scaled simulated space gaps $\iota = g_s / \langle g_s \rangle$ and
  (b) un-scaled simulated space gaps $g_s$ between successive cars in traffic flow.}
\label{fig:5}       
\end{figure}

Fig.~\ref{fig:5} shows the corresponding distributions of the
simulated space gaps under different traffic pressures. As the
traffic flow density (occupancy) increases, the distribution of
space gaps will become more concentrated, which is in accordance
with the practical observations (see the un-scaled distributions
shown in Fig.~3a in \cite{HelbingTreiberKesting2006} for
comparison). Moreover, the shape of the un-scaled simulated gap
distributions look similar to the empirical data and also the scaled
distributions predicted by using random matrix theory (see that
shown in Fig.~1 in \cite{Abul-Magd2007} for comparison).

%

\begin{figure}[h]
\center
  \resizebox{0.6\columnwidth}{!}{%
  \includegraphics{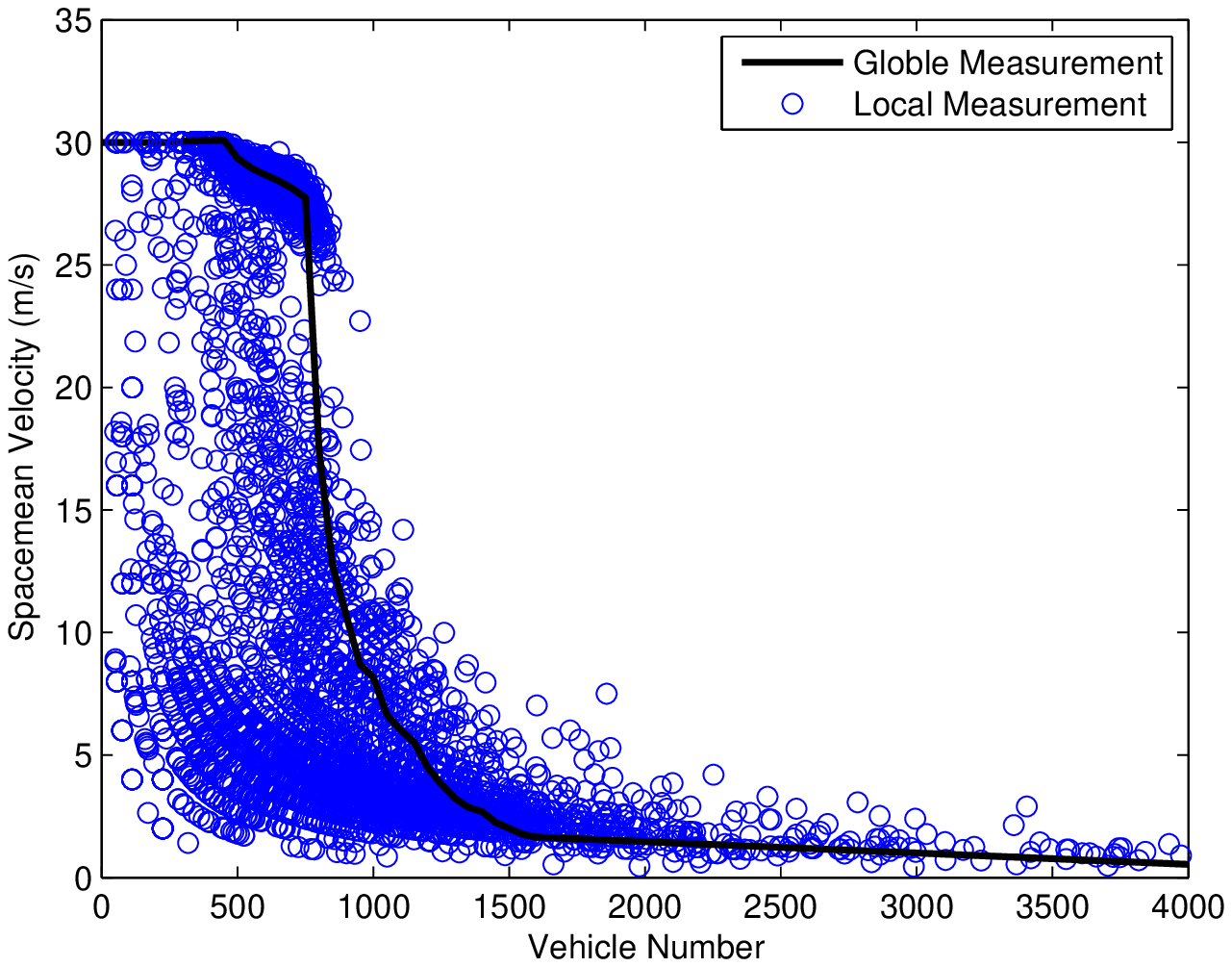}
  }
  \centerline{\footnotesize (a)}
  \resizebox{0.6\columnwidth}{!}{%
    \includegraphics{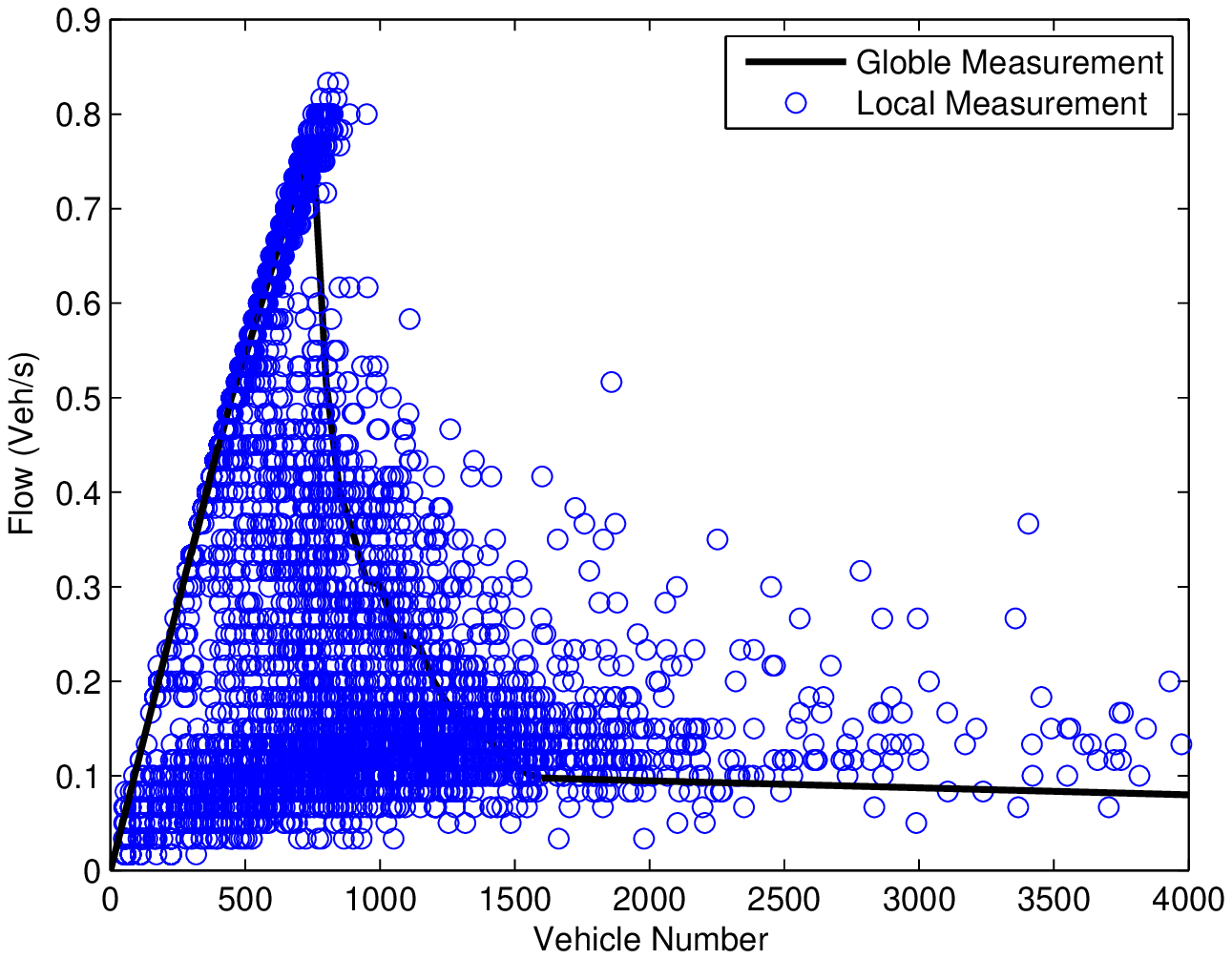}
  }
  \centerline{\footnotesize (b)}
\caption{(a) The $(k, \bar{v}_s)$ diagram for the proposed model,
obtained by local and global measurements. (b) The $(k, q)$ diagram
for the proposed model, obtained by local and global measurements.}
\label{fig:6}       
\end{figure}

Fig.~\ref{fig:6} show the $(k, \bar{v}_s)$ and $(k, q)$ diagrams for
the proposed model, obtained by local and global measurements. These
macroscopic traffic stream characteristics are measured according to
\cite{MaerivoetMoor2005}. Here, $K_g$ denotes the length of the
closed single-lane system, which is $27000$m. It can be found that
the local measurements in Fig.~\ref{fig:6} also discriminate
Kerner's three-phase flows: the free-flow regime contains only a few
data points on a line; the synchronized regime is formulated by
widely scatters of the data points; and jammed regimes contains the
data points corresponding to Kerner's line $J$.

%

\begin{figure}[h]
\center
  \resizebox{0.4\columnwidth}{!}{%
  \includegraphics{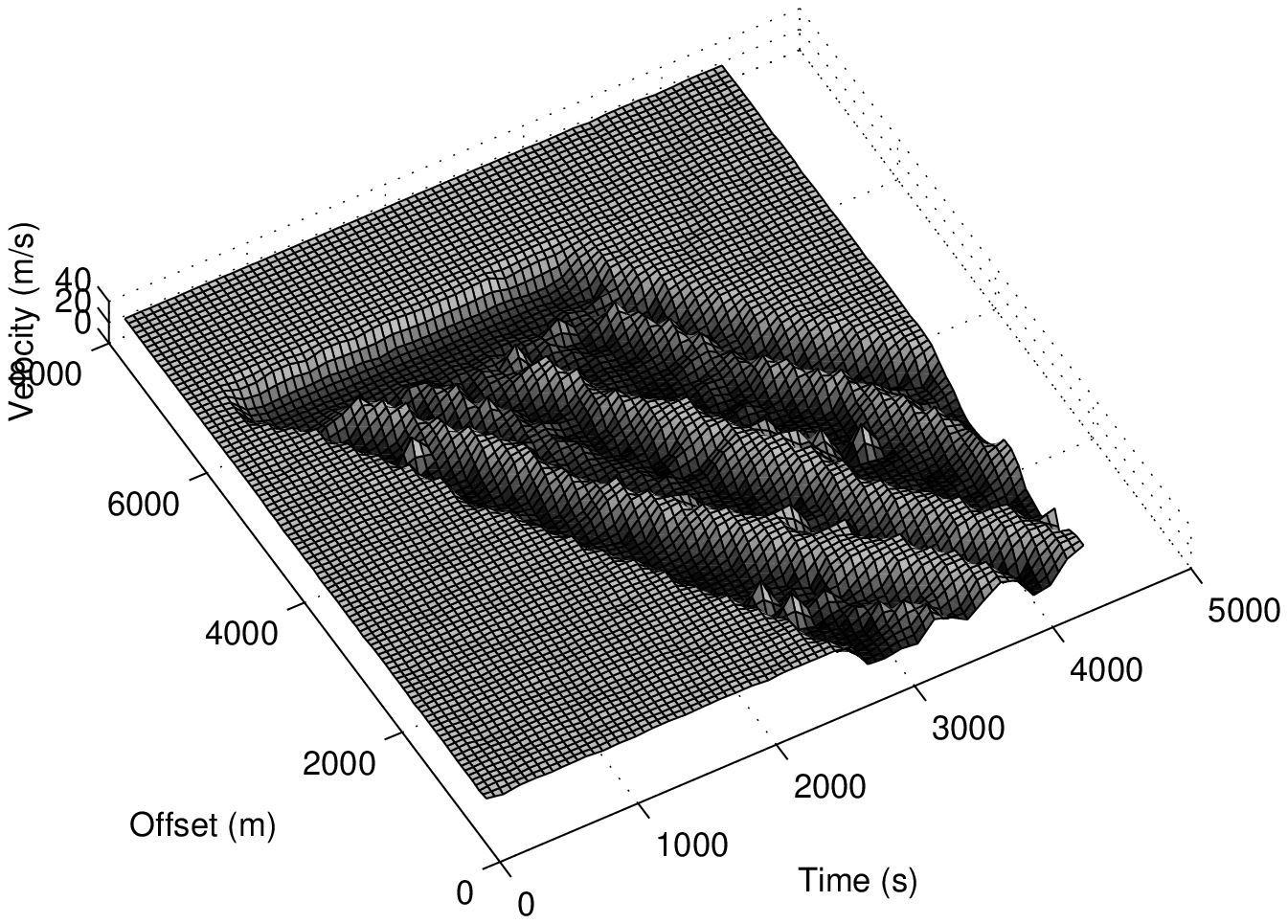}
  }
  \centerline{\footnotesize (a)}
  \resizebox{0.4\columnwidth}{!}{%
    \includegraphics{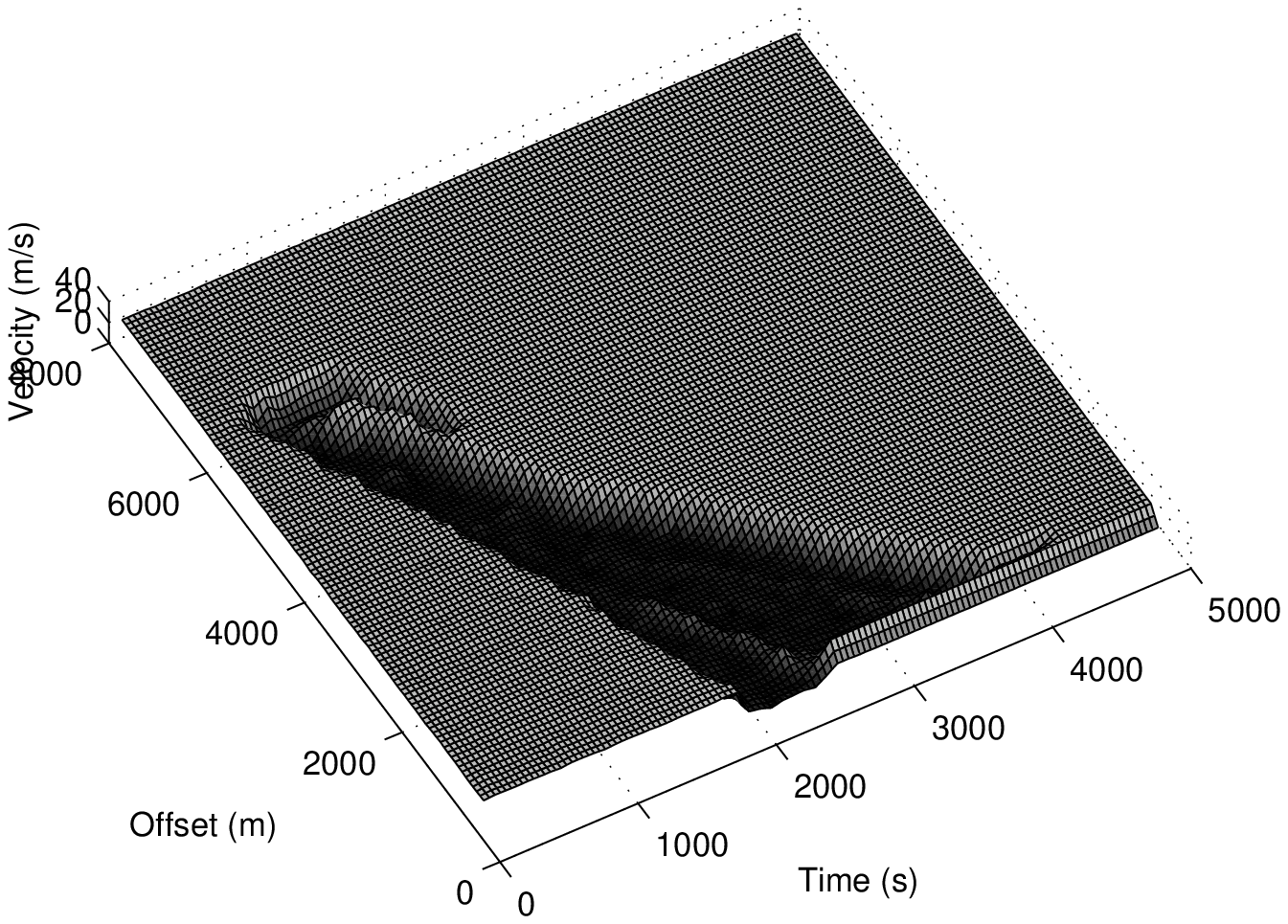}
  }
  \centerline{\footnotesize (b)}
  \resizebox{0.4\columnwidth}{!}{%
    \includegraphics{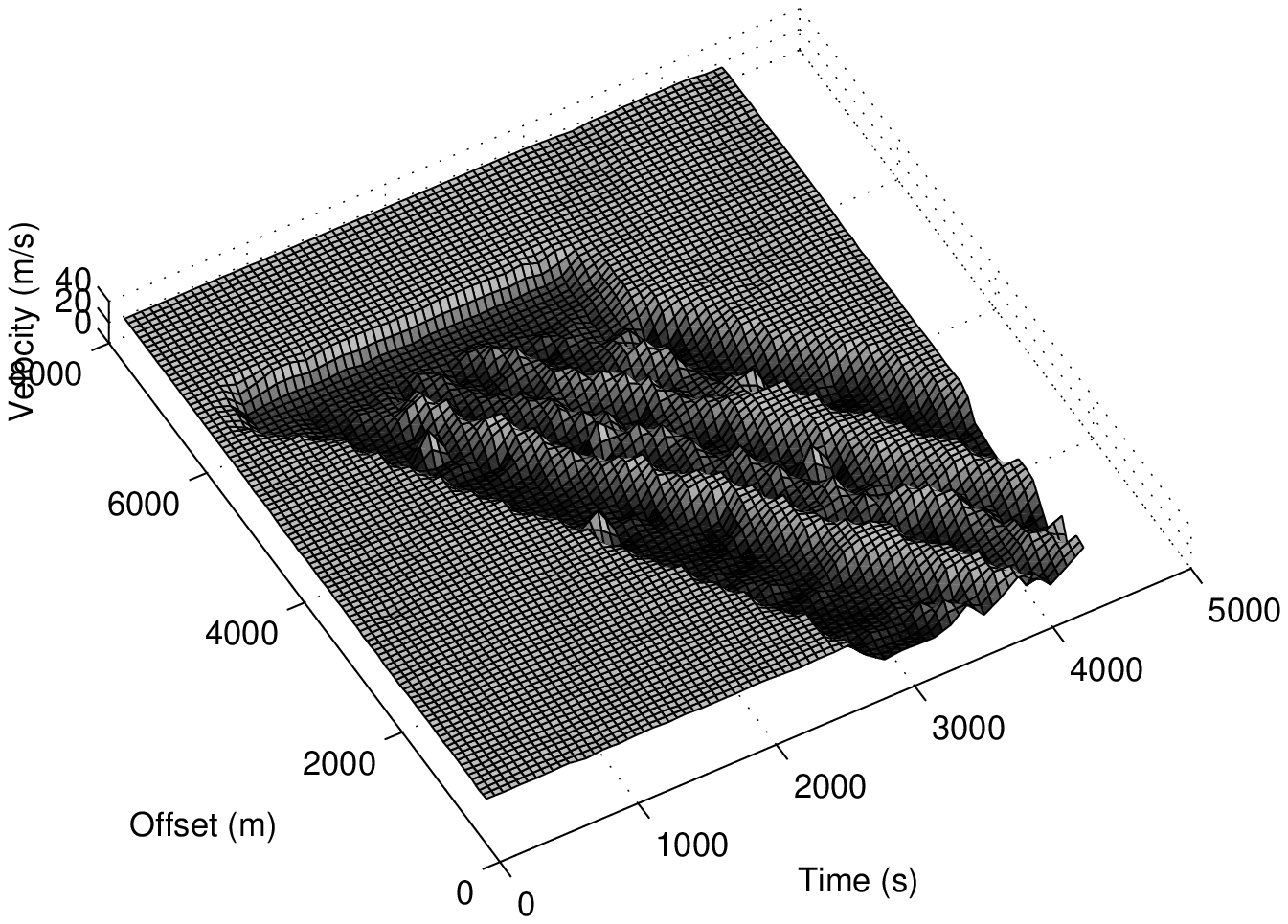}
  }
  \centerline{\footnotesize (c)}
\caption{The spatiotemporal speed diagrams of: (a) the oscillation
in traffic flow; (b) the wide moving jam; and (c) the stop-and-go
waves.}
\label{fig:7}       
\end{figure}

Fig.~\ref{fig:7} shows three typical complex phenomena reproduced by
using the proposed model: (a) the oscillation in traffic flow; (b)
the wide moving jam; and (c) the stop-and-go waves. This proves that
the proposed model is capable to describe the other complex dynamic
features of road traffic flow, too.

%

\appendix*

\section{Explanation of the Log-Normality}
\label{sec:4}

According to the Galton board model, if the variation dynamics of
time headway should be depicted by the following equation without
losing any generality.

\begin{equation}
\label{equ:6} t_h(t) = \left\{
\begin{array}{ll} \beta t_h(t-T), \indent \textrm{with probability $(1-p)$}\\
\frac{1}{\beta} t_h(t-T), \indent \textrm{with probability $p$}
\end{array} \right.
\end{equation}

Notice that be definition, we roughly have

\begin{equation}
\label{equ:7} t_h(t) \approx \frac{g_s(t)}{v(t+T)}
\end{equation}

we can then have

\begin{equation}
\label{equ:8} \frac{g_s(t)}{v(t+T)} = \left\{
\begin{array}{ll} \beta \frac{g_s(t-T)}{v(t)}, \indent \textrm{with probability $(1-p)$}\\
\frac{1}{\beta} \frac{g_s(t-T)}{v(t)}, \indent \textrm{with
probability $p$}
\end{array} \right.
\end{equation}

Considering the pre-determined limits of the velocity and
ac/deceleration rates, we can directly derive Eq.(\ref{equ:4}) from
Eq.(\ref{equ:8}) then.

\begin{acknowledgments}

This work was supported in part by National Basic Research Program
of China (973 Project) 2006CB705506, Hi-Tech Research and
Development Program of China (863 Project) 2006AA11Z208,
2006AA11Z229 2007AA11Z222, and National Natural Science Foundation
of China 50708055.

\end{acknowledgments}

%
%

\end{document}